\documentclass[10pt,letterpaper]{article}
\usepackage{opex3}
\usepackage{subfigure}

\begin{document}

\title{Titanium nitride as a plasmonic material for visible wavelengths}

\author{Gururaj V. Naik$^1$, Jeremy L. Schroeder$^2$, Timothy D. Sands$^{1,2}$ and Alexandra Boltasseva$^{1,3,*}$}

\address{$^1$School of Electrical \& Computer Engineering and Birck Nanotechnology Center, Purdue University Indiana 47907 USA\\
$^2$School of Materials Engineering and Birck Nanotechnology Center, Purdue University Indiana 47907 USA\\
$^3$DTU Fotonik, Technical University of Denmark, Lyngby 2800, Denmark}

\email{aeb@purdue.edu} 



\begin{abstract}
The search for alternative plasmonic materials with improved optical properties over the traditional materials like silver and gold could ultimately lead to the long-awaited, real-life applications of plasmonic devices and metamaterials. In this work, we show that titanium nitride could perform as an alternative plasmonic material in the visible wavelength range. We evaluate the performances of various plasmonic and metamaterial structures with titanium nitride as a plasmonic component and show that transformation-optics devices and an important class of metamaterials with hyperbolic dispersion based on TiN could significantly outperform similar devices based on conventional metals in the visible wavelengths range.
\end{abstract}




\section{Introduction}
The rapidly advancing fields of plasmonics \cite{review_barnes,LSPR_review} and relatively new areas of metamaterials (MMs) and transformation optics (TO) \cite{TO_pendry,NIMreview_soukoulis,reviewSci_shalaev} continue to boom with revolutionary ideas and exciting experimental realizations. Fantastic ideas such as negative refractive index materials \cite{perfectlens_pendry}, nanoscale imaging \cite{hyperlens_jacob,superlens_ramakrishna}, invisibility cloaks \cite{cloak_cai} and light concentrators \cite{TO_kildishev,OBH_narimanov} have paved the way for the realization of devices with unprecedented functionalities \cite{superlens_zhang,hyperlens_zhang,NIM_shalaev,NIM_wegener,1st_cloak_smith,3Dcloak_wegener}. Although the experimental realizations of these devices serve as proofs-of-concepts, their practical applications are still out of reach. One of the major challenges is the large loss coming from the plasmonic components of these devices. Even metals with the highest conductivities (e.g., silver and gold) exhibit excessive losses at optical frequencies \cite{optprop_JC}. Another major issue associated with the use of conventional metals is that their real part of permittivity is too large in magnitude to be useful for many TO and MM devices \cite{TO_kildishev}.\\

In the search for better plasmonic materials, transparent conducting oxides (TCOs) were proposed as low loss alternatives to gold and silver in the near-infrared \cite{APM_LPR,AZO_RRL}. However, TCOs cannot be plasmonic at visible wavelengths because their carrier concentration is limited to around $10^{21}cm^{-3}$ \cite{reviewTCO_minami}. In this paper, we show that titanium nitride could be a good candidate for an alternative plasmonic material at visible wavelengths. We first discuss the processing details and optimizations involved in the deposition of titanium nitride films. We then address the performance of various plasmonic and metamaterial devices based on titanium nitride through the figures-of-merit \cite{FOM_berini,APM_LPR}.

\section{Titanium nitride: Deposition and characterization}
Transition metal nitrides like TiN are ceramic materials that are non-stoichiometric. Their composition and hence, the optical properties depend significantly on the preparation method. Some of these nitrides possess metallic properties at visible wavelengths because of large free carrier concentration ($\approx 10^{22} cm^{-3}$) \cite{nitrides_hiltunen,TaN_aouadi,TiN_patsalas}. However, high interband losses make many of these compounds unattractive for plasmonic applications. Titanium nitride, however, exhibits smaller interband losses in the red part of the visible spectrum and a small negative real permittivity \cite{TiN_patsalas}. This makes titanium nitride a promising material for plasmonic metamaterial applications in the visible spectrum.\\

\begin{figure}[tb]
\centering
\mbox{\subfigure{\includegraphics[width=6.66cm]{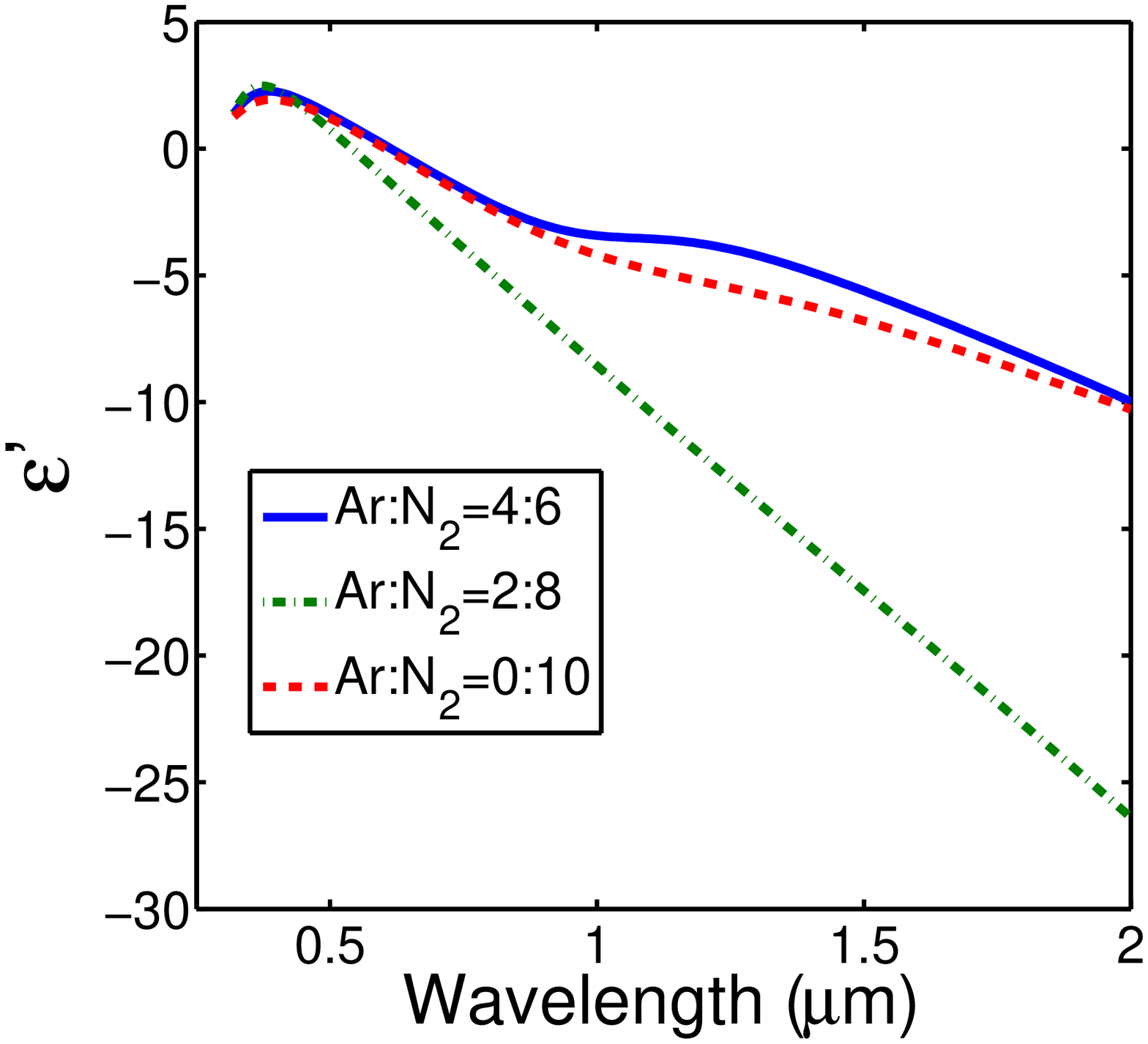}
\quad
\subfigure{\includegraphics[width=6.66cm]{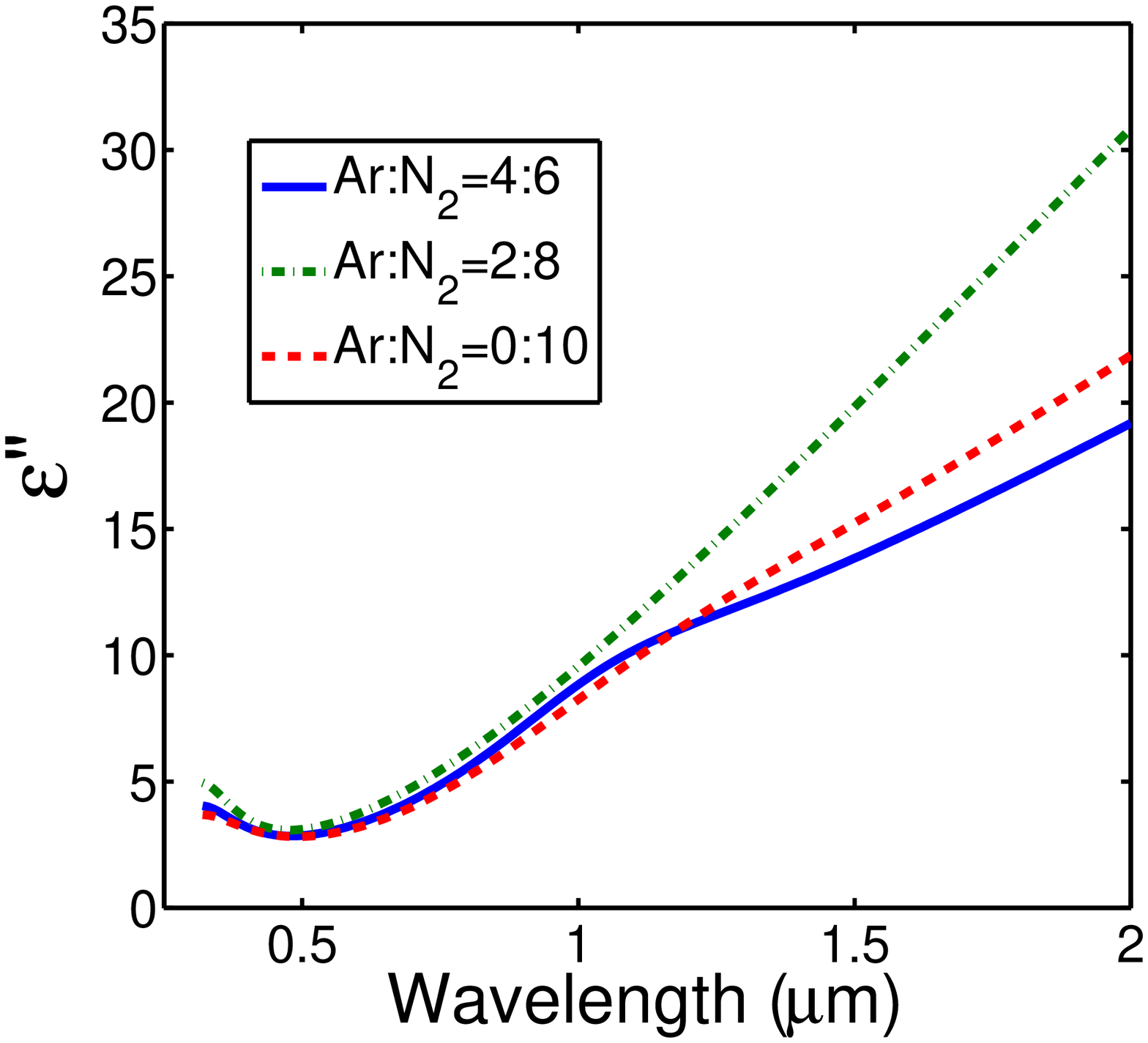} }}}
\caption{Dielectric function of sputter-deposited titanium nitride films retrieved by ellipsometry measurements. The films were deposited on on c-sapphire substrates by DC reactive magnetron sputtering at 300$^oC$ with different flow ratios (sccm:sccm) of argon and nitrogen.}
\label{fig1}
\end{figure}

Thin films of titanium nitride were deposited on c-sapphire substrates by DC reactive magnetron sputtering (PVD Products Inc.) of a 99.995\% titanium target in an argon-nitrogen environment. The base pressure before deposition was $4\times10^{-8}$ Torr. The films were deposited at a pressure of 5 mTorr with varying flow ratios of argon and nitrogen (Ar (sccm):$\mbox{N}_2$ (sccm) of 4:6, 2:8 and 0:10). The sputtering power was held constant for all depositions at 200 W (DC) and the target-substrate distance was 8 cm. The deposition rate was approximately 25 $\AA$/min and about 150 nm thick films were deposited. The substrate temperature during deposition was held at 300$^oC$ or 500$^oC$. The resulting films were characterized by a variable angle spectroscopic ellipsometer in the near-IR, visible and near-UV wavelength ranges. A Drude-Lorentz (with 3 Lorentz oscillators) model provided good fit to the ellipsometric measurements. Figure \ref{fig1} shows the dielectric function retrieved for three TiN films deposited at 300$^oC$ with different flow ratios of argon and nitrogen. Drude damping of the carriers does not change significantly among the three films. However, the unscreened plasma frequency increases slightly for the film deposited at flow ratio of 2:8 (Ar: $\mbox{N}_2$), which results in higher loss compared to the other films.\\

\begin{figure}[tb]
\centering
\mbox{\subfigure{\includegraphics[width=6.66cm]{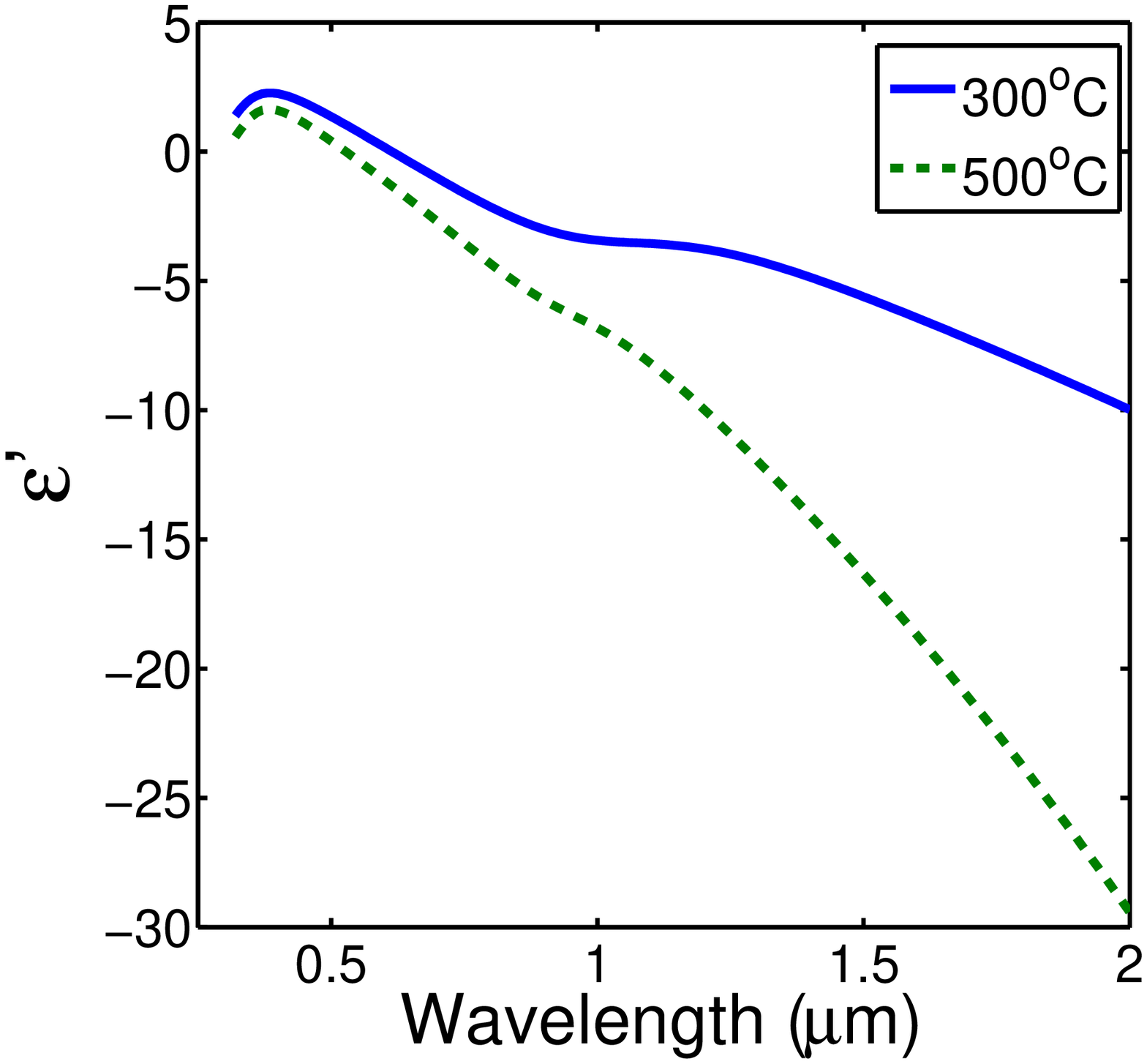}
\quad
\subfigure{\includegraphics[width=6.66cm]{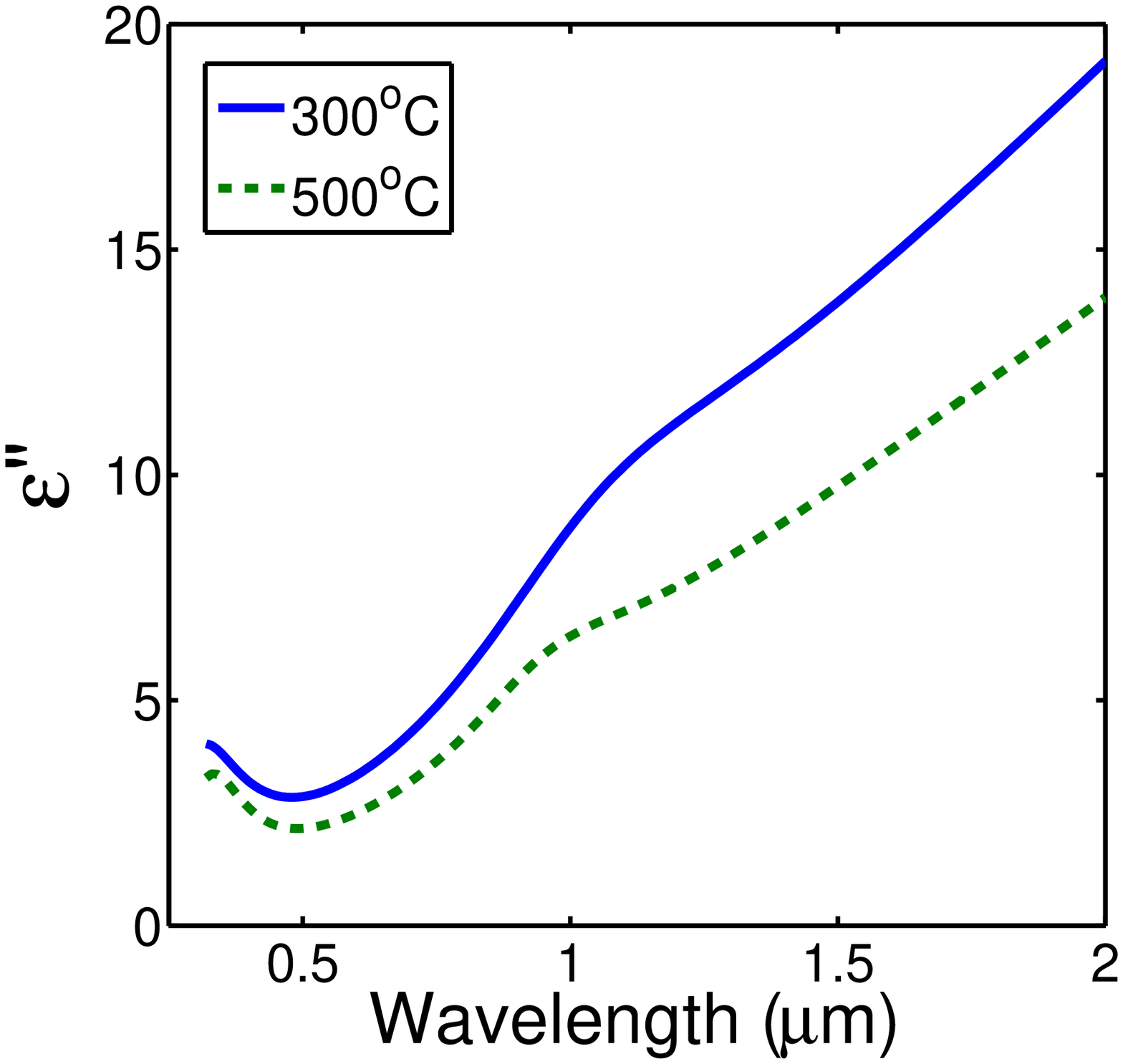} }}}
\caption{Dielectric function of titanium nitride films deposited at 300$^oC$ and 500$^oC$ retrieved by ellipsometry measurements. The films were deposited with the flow ratio of argon and nitrogen set to 4:6.}
\label{fig2}
\end{figure}

The influence of substrate temperature on the optical properties of these films was studied by depositing TiN films at substrate temperatures of 300$^oC$ and 500$^oC$ with the Ar:$\mbox{N}_2$ flow ratio fixed at 4:6. Figure \ref{fig2} shows the dielectric function extracted for these two films. Clearly, the film deposited at the higher temperature shows lower loss. In the rest of the discussions on figures-of-merit, the optical properties of a TiN film deposited at 500$^oC$ are used.

\section{Plasmonic metamaterial applications}
In this section, we consider TiN as a plasmonic building block for different applications: surface-plasmon-polariton (SPP) waveguides, localized surface plasmon resonance (LSPR), hyperbolic metamaterials and general transformation optics devices. The figures-of-merit of these devices with TiN as plasmonic component are evaluated and compared against their metal based counterparts.\\

\subsection{Plasmonic applications}
\begin{figure}[htb]
\centering
\mbox{\subfigure{\includegraphics[width=6.66cm]{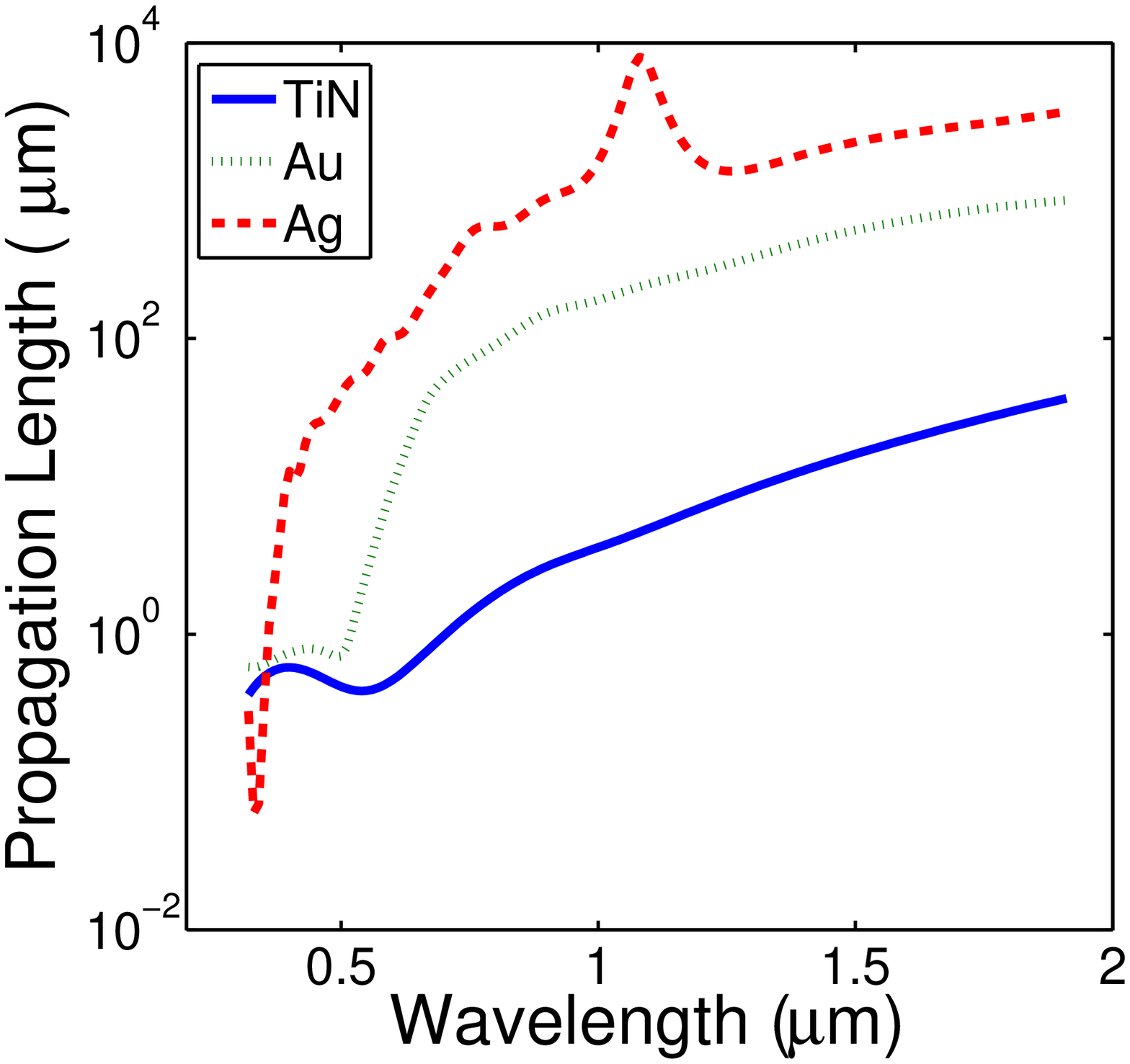}
\quad
\subfigure{\includegraphics[width=6.66cm]{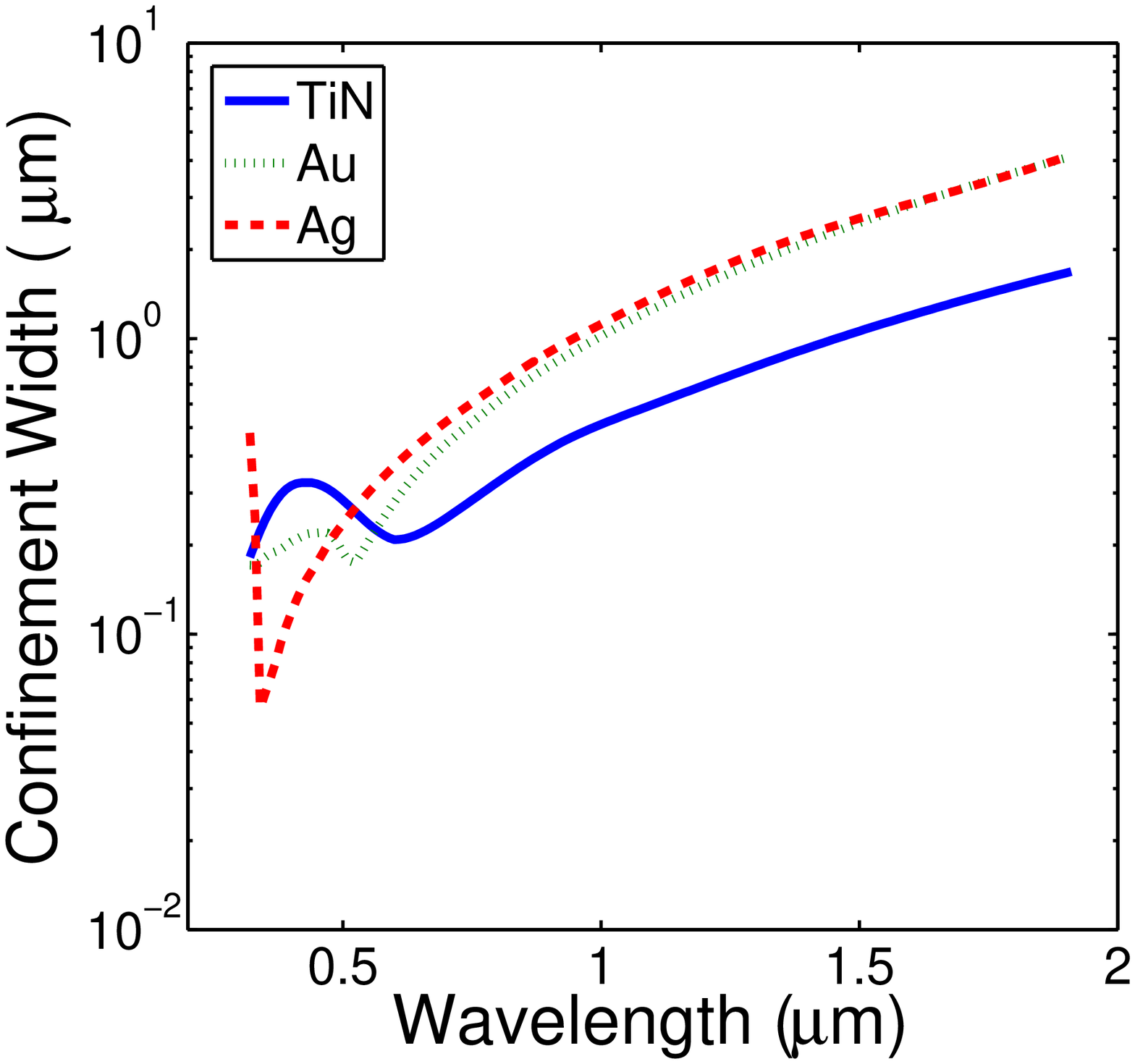} }}}
\caption{Comparison of the performance characteristics of SPP waveguides formed by TiN-, gold- and silver- air interfaces: a) Propagation length (1/e field decay length along the propagation direction) b) Confinement width (1/e field decay widths on each side of the interface).}
\label{fig3}
\end{figure}

\begin{figure}[htb]
\centering
\includegraphics[width=6.66cm]{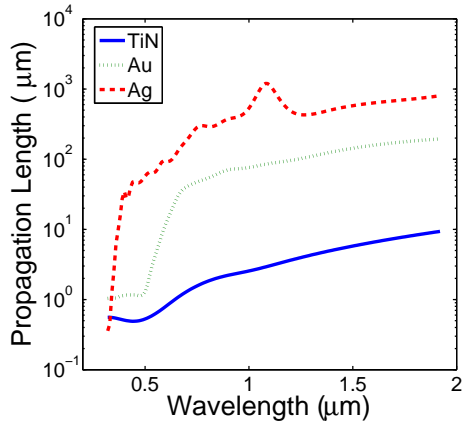}
\caption{Propagation length of the lowest order long-range SPP mode in a metal/300 nm air gap/metal structure where the metal is TiN, gold or silver.}
\label{fig4}
\end{figure}

An SPP propagating along a single metal/air interface is characterized by two important parameters: propagation length (defined as the 1/e field-decay length along the direction of propagation) and confinement width (the 1/e field-decay width on each side of interface) \cite{FOM_berini}. There is often a trade-off between these parameters such that longer propagation lengths require larger confinement widths and smaller confinement widths result in smaller propagation lengths. Figure \ref{fig3} shows the propagation length and confinement of an SPP along TiN-, gold- and silver-air interfaces. We note that TiN gives slightly better confinement than either traditional plasmonic metal. However, the propagation length for TiN is an order of magnitude smaller than for either gold or silver. Thus, TiN exhibits no better trade-off between the propagation length and the confinement of SPPs than what can be obtained with conventional metals. This fact can be appreciated more directly when a metal/air/metal SPP waveguide is considered. If the air gap is 300 nm (which is equal to the confinement width), the propagation length of the lowest-order long-range SPP mode is about an order of magnitude larger in the case of conventional metals than for the TiN case (see Fig. \ref{fig4}).\\

\begin{figure}[htb]
\centering
\includegraphics[width=6.66cm]{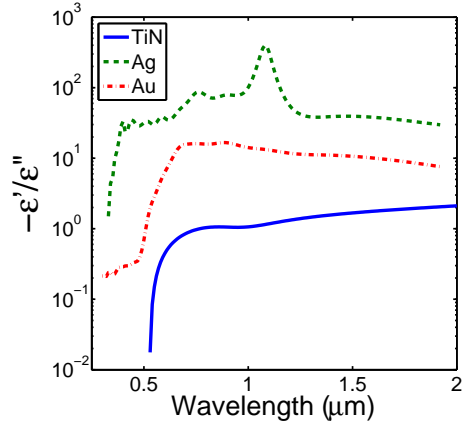}
\caption{Figure-of-merit for LSPR applications with spherical nanoparticles of TiN, gold or silver.}
\label{fig5}
\end{figure}

While different from SPP modes, localized surface plasmon resonance or LSPR modes are useful in many sensing applications involving local field enhancement \cite{LSPR_review}. Plasmonic metallic nanoparticles can support LSPR modes and enable applications. The figure-of-merit for spherical metallic nanoparticles can be approximated by $-\epsilon'/\epsilon"$ \cite{APM_LPR}. The LSPR figure-of-merit for conventional metals and TiN nanoparticles is plotted in Fig. \ref{fig5}. Although the conventional metals dominate in the visible spectrum, TiN can be a good substitute for metals at longer wavelengths in the mid-infrared.

\subsection{Hyperbolic Metamaterials and transformation optics}
\begin{figure}[htb]
\centering
\includegraphics[width=6.66cm]{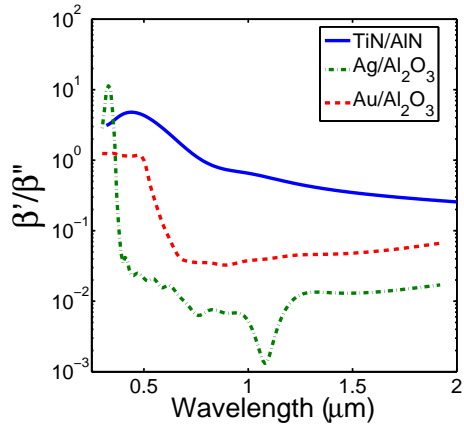}
\caption{Figure-of-merit of HMMs formed by alternating, sub-wavelength layers of metal/dielectric (TiN/AlN, silver/alumina and gold/alumina).}
\label{fig6}
\end{figure}

Metamaterials with hyperbolic dispersion are receiving increased attention from researchers because of their unique properties such as the propagation of extremely high-k waves \cite{hyperlens_jacob} and a broadband singularity in the photonic-density-of-states (PDOS) \cite{BBpurcell_jacob}. These properties have resulted in devices such as the hyperlens \cite{hyperlens_jacob} and phenomena such as engineering the spontaneous emission rate \cite{PDOSengg_jacob}, both of which have useful applications in quantum optics. Hyperbolic metamaterials (HMMs) can be easily fabricated by stacking alternating, sub-wavelength layers of metal and dielectric materials. For example, the first demonstration of a hyperlens used alternating layers of silver and alumina \cite{hyperlens_zhang}. However, such a material system works well only in the near-UV where the performance of silver is best-suited for HMMs. In the visible spectrum, neither gold nor silver can produce high performance HMMs. To compare the performances, we adopt the figure-of-merit from \cite{negref_hoffman} as $Re\{\beta_{\perp}\}/Im\{\beta_{\perp}\}$ where $\beta_{\perp}$ is the propagation vector of an electromagnetic wave along the direction perpendicular to the plane of the metal/dielectric layers. Figure \ref{fig6} shows the figure-of-merit of HMMs formed by alternating layers of silver/alumina, gold/alumina and TiN/AlN. Clearly, the TiN system outperforms the metal based systems in the red part of the visible spectrum. For wavelengths shorter than 500 nm, TiN is not plasmonic; therefore, the figure-of-merit values are valid only for wavelengths longer than 500 nm.

\begin{figure}[htb]
\centering
\mbox{\subfigure{\includegraphics[width=6.66cm]{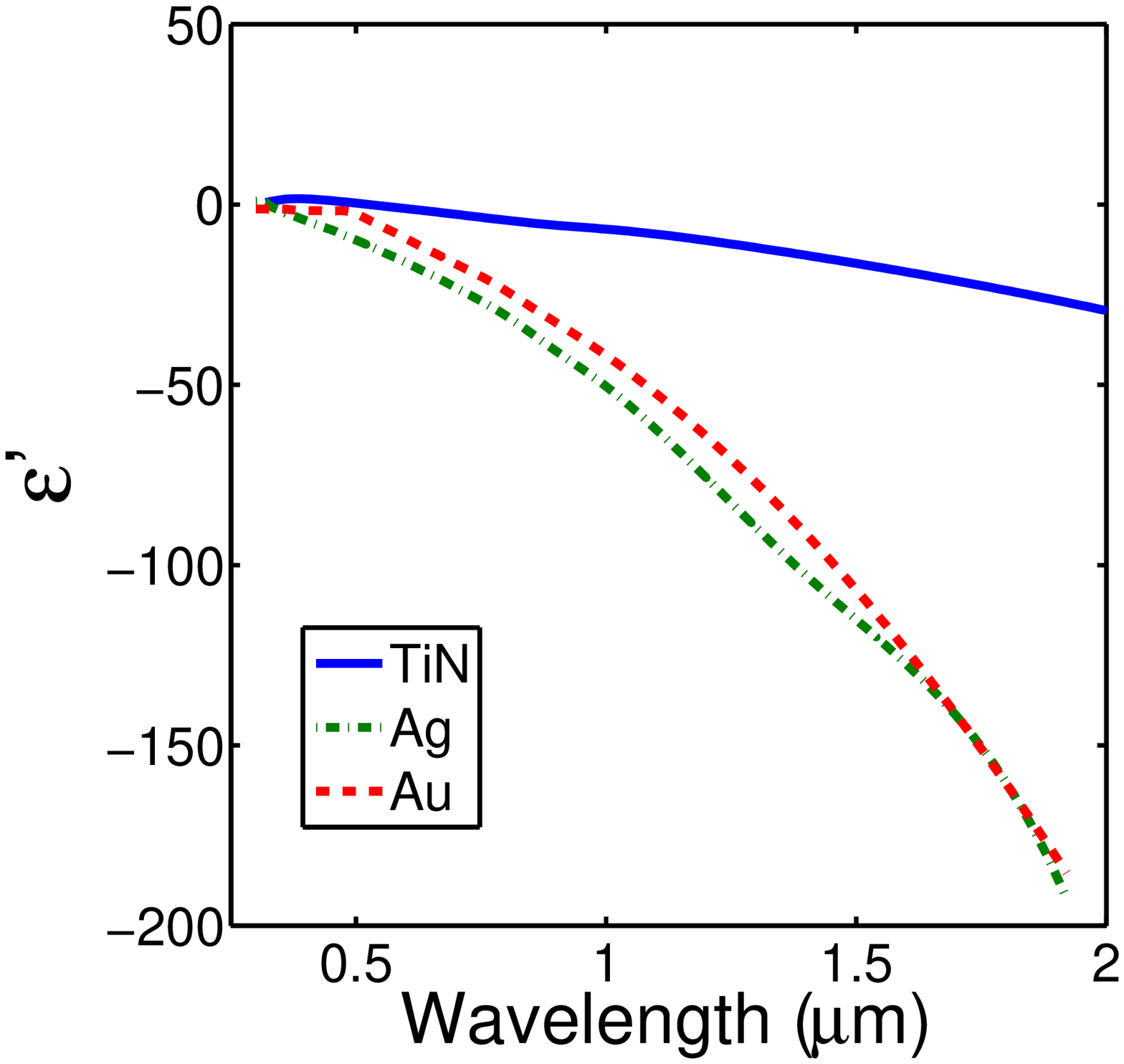}
\quad
\subfigure{\includegraphics[width=6.66cm]{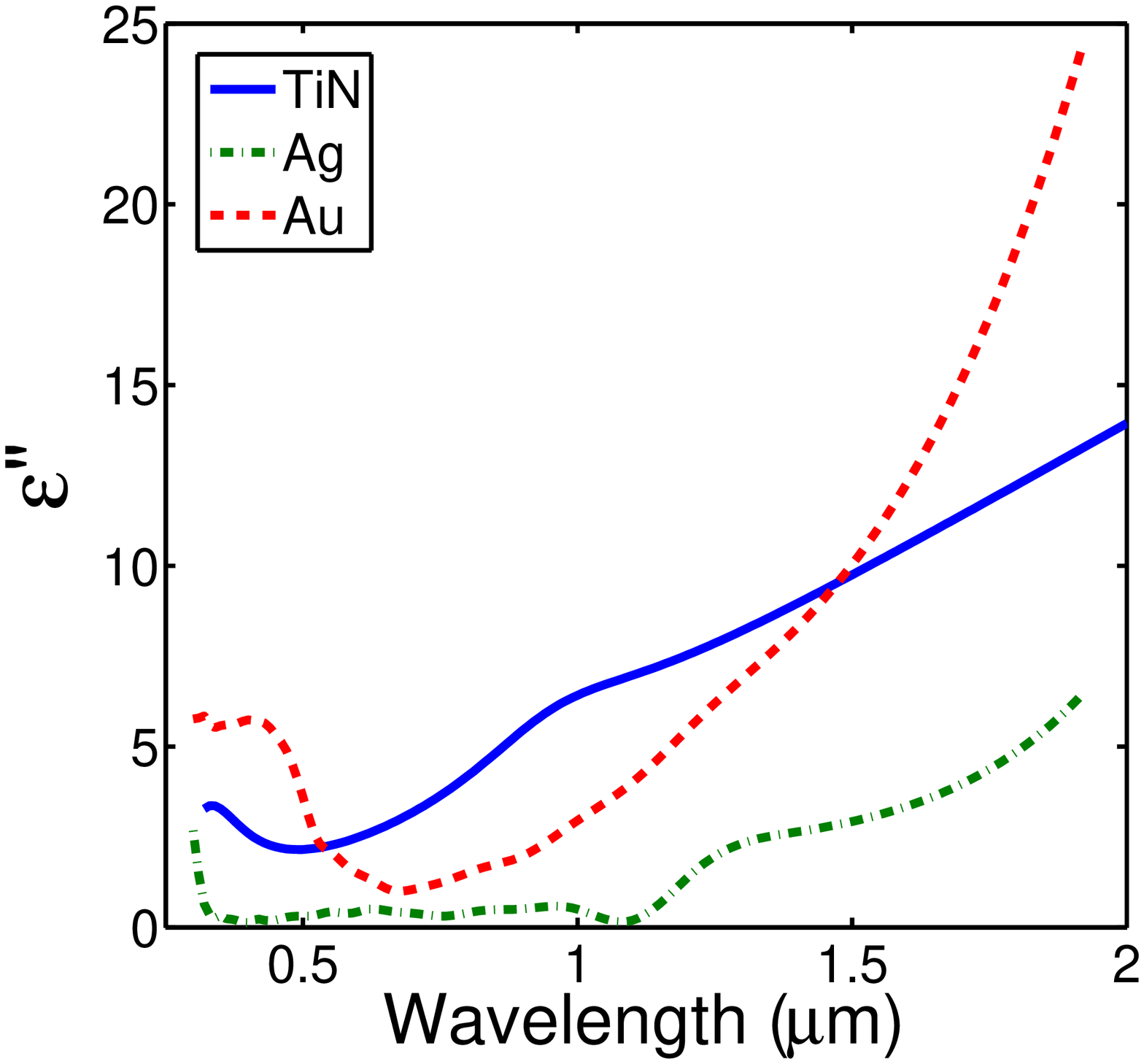} }}}
\caption{Dielectric function of TiN in comparison with conventional plasmonic materials (gold and silver).}
\label{fig7}
\end{figure}

In general, transformation optics requires plasmonic components whose real permittivity values are on the order of unity. While none of the conventional metals satisfy this condition, titanium nitride does meet this criterion and is therefore a suitable material that could enable transformation optics in the visible range. As a comparison, the dielectric functions of TiN and bulk, conventional metals are plotted in Fig. \ref{fig7}. The figure clearly shows the disadvantage of conventional metals in terms of the real part of permittivity. However, the imaginary part of permittivity, which signifies the losses in the material, is the lowest in the case of silver. TiN is better than gold  only for longer wavelengths. In practical applications, it is rather difficult to obtain low, bulk-like losses in designs using silver because of problems such as surface roughness, grain boundary scattering and oxidation. On the contrary, TiN does not possess such difficulties and forms a better choice than conventional metals for applications such as transformation optics.

\section{Conclusion}
We show that titanium nitride can serve as an alternative plasmonic material for plasmonic and metamaterial applications in the visible frequencies. From the comparative study of the performance of various devices, we conclude that the hyperbolic metamaterials and transformation optics devices with titanium nitride as a plasmonic building block significantly outperform their metal-based counterparts. Thus, titanium nitride is a better plasmonic material for metamaterial and transformation optics in the visible wavelengths.

\section*{Acknowledgments}
We thank Prof. Vladimir M. Shalaev for the useful discussions. This work was supported in part by ONR MURI on "Large-Area 3D Optical Metamaterials with Tunability and Low Loss."\\

\bibliography{final}

\end{document}